\documentstyle[preprint,aps]{revtex}
\begin{document}
\draft
%\preprint{dvi file made on \today}

\title
{Suppression of the Kondo Effect in Quantum Dots by Even-Odd Asymmetry}
\author{Yi Wan$^1$, Philip Phillips$^1$ and Qiming Li$^2$}
\vspace{.05in}

%
%\begin{instit}
\address
{$^1$Loomis Laboratory of Physics\\
University of Illinois at Urbana-Champaign\\
1100 W.Green St., Urbana, IL, 61801-3080}
\address
{$^2$Ames Laboratory and Department of Physics\\
Iowa State University\\
Ames, Iowa  50011}
%\end{instit}
%
\maketitle

\begin{abstract}
We analyze here a model for single-electron charging in semiconductor quantum
dots that includes the
standard Anderson on-site repulsion (U) as well as the spin-exchange ($J_d$)
that is
inherently present among the electrons occupying the various quantum levels of
the
dot.  We show explicitly that for ferromagnetic
coupling ($J_d>0$), an s-d exchange for an S=1 Kondo problem is recovered.  In
contrast,
for the antiferromagnetic case, $J_d<0$, we find that the Kondo
effect is present only if there are an odd number of electrons on the dot.
In addition, we find that spin-exchange produces a second
period in the conductance that is consistent with experimental measurements.
\end{abstract}
\vspace{.1in}

\pacs{PACS numbers:}

\narrowtext

When a gate voltage is applied to a nano-scale semiconductor inversion layer
(or quantum dot), electrons
will flow one at a time across this device provided that the applied voltage is
an integral multiple
of the capacitance charging energy of the quantum dot.  Experiments
illustrating
the principle of charge quantization by virtue of the charging energy have been
performed recently on numerous
semiconductor\cite{likharev}\cite{kastner}\cite{ashoori}
\cite{adesida}\cite{beenakker} as well as superconducting
\cite{tinkham} nano-structures.  We focus here solely on the semiconductor
devices.
It is now well-accepted\cite{beenakker} that in semiconductor quantum dots, the
dominant contribution
to the capacitance charging energy, $E_c=\frac {e^2} {2C}$ arises from the
on-site
Coulomb repulsion.  Here C is the capacitance between the quantum dot, the
tunnel junctions, and the
electrical leads connected to the dot.  Transport in quantum dots will be
Coulomb limited
if $k_BT<E_C$ and $k_BT> \Delta\epsilon$, where $\Delta\epsilon$ is the spacing
between the
single particle states of the dot.

Because of the central role played by on-site Coulomb repulsions in the
transport properties
of quantum dots, it is natural to model a quantum dot with a Hubbard-like
model.
In so far as a quantum dot can be reduced to a single site \cite{lee}
with a charging energy U, the Anderson model\cite{anderson} for the interaction
of a magnetic defect coupled
to a non-interacting sea of conduction electrons is appropriate\cite{lee}:
\begin{eqnarray}
H_A= \sum_{k,\sigma}
\epsilon_{k} a_{k\sigma}^{\dagger} a_{k\sigma} + \sum_{\sigma} \epsilon_{d}
a_{d\sigma}^{\dagger} a_{d\sigma} + \sum_{k, \sigma} V_{kd}
(a_{k\sigma}^{\dagger} a_{d\sigma}
+a_{d\sigma}^{\dagger} a_{k\sigma}) + U n_{d\uparrow} n_{d\downarrow}\\
= H_0 +  \sum_{k, \sigma} V_{kd} (a_{k\sigma}^{\dagger} a_{d\sigma}
+a_{d\sigma}^{\dagger} a_{k\sigma})
\end{eqnarray}
where $\epsilon_d$ is the defect energy of the magnetic impurity, $V_{kd}$ the
overlap integral between a band state with momentum k and the impurity,
$a_{k}^{\dagger}$ creates an electron in the band states,
$a_{d\sigma}^{\dagger}$ creates an electron with spin $\sigma$ on the
impurity, and $n_{d\sigma} = a_{d\sigma}^{\dagger}a_{d\sigma}$ is the
number operator for an electron of spin $\sigma$.  As a consequence of the
on-site
repulsion, the single particle states on the impurity have energies,
$\epsilon_d$ and
$\epsilon_d+U$.   At high temperatures, the
density of states of this model has two Lorentzian peaks centered at these two
energy
levels.  At low temperatures,
however, the Anderson model displays a Kondo resonance\cite{swolff}\cite{amw}
at the Fermi level.
Although the Kondo resonance is expected to occur for any value of the defect
energy
within the range $-U<\epsilon_d<0$, it is most favourable at the defect energy
corresponding
to the greatest stability of the local moment at the d-impurity, namely,
$\epsilon_d=-\frac {U} {2}$. At this energy
$H_A$ is particle-hole symmetric, and the Kondo resonance is pinned at
$\epsilon_F=0$.
The single-particle states $\epsilon_d$ and
$\epsilon_d+U$ lie symetrically, then, around the Fermi energy.  When
the chemical potential of the source lead coincides with the energy of the
Kondo resonance, a single electron should charge the
dot. In the symmetric limit, this state of affairs should obtain at
half-integer multiples of the  charging energy.
Thus far, no experimental hint of the Kondo resonance has been observed in
quantum dots in zero bias voltage.

It is precisely the conditions under which the Kondo effect should be observed
in quantum
dots that we address here. While it may be premature to draw any conclusion
from the
lack of experimental confirmation of the Kondo effect, it is certainly
appropriate
to investigate the validity of the Anderson model to a quantum dot.  It is in
addressing
this issue that we are 1) able to predict even-odd charging effects in
semiconductor quantum
dots as well as 2) a suppression of the Kondo effect when a quantum dot has an
even number of electrons.  The most obvious inadequacy of the Anderson model in
the context
of quantum dots is the truncation of the multiple electronic levels on a
quantum dot
to a single state.  If multiple electronic levels are included on the dot,
then other Coulomb interactions besides the on-site U become relevant. A key
quantity
that comes into play is the intrinsic Coulomb exchange energy, $J_d$ between
two levels. Consider for the moment a two-orbital
model for a quantum dot.   For two degenerate levels and
$U\approx$ the direct Coulomb exchange integral, the energy of the two-body
states predicted by this model
are $2\epsilon_d+U-\frac {J_d} {4}$, $2\epsilon_d+U$, and
$2\epsilon_d+U+\frac {3J_d} {4}$. The energy of the 3-electron state is
$3(\epsilon_d+U)$.
Consequently, the charging energy depends on $J_d$.  In fact,
the general role of spin-exchange is to introduce a spin-dependent charging
energy
that is determined by the parity of the total number of electrons on the dot.
It is
worth pointing out that charging experiments on controlled-barrier
atoms\cite{kastner} in strong magnetic fields display
 systematic oscillation in the peak heights,
widths as well as in the separations that are consistent with a second period
in the
conductance as a function of the applied gate voltage. Such systematic
deviations
have been attributed to a splitting of the energy between the up and down
Landau
 levels, rather than to spin exchange.  We propose here that such trends are
also
consistent with a spin-exchange model.

We explore then the simplest model of a quantum dot
that includes the effects of spin-exchange.  A natural way of including spin
exchange
is simply to introduce another level into the Anderson model.
The only qualitative change this level is going to provide is the
spin-interaction with
the d-level.  Consequently, we treat this level as a local spin, $\bf S$.
Because
spin-exchange plays no role if $S=0$, we will consider only the case
in which the S-level is singly occupied, or equivalently, $S^z=\pm\frac {1}
{2}$.  Hence, large N expansion techniques are
inappropriate to solving this problem. If we label the spin on the d-level of
the impurity with $\bf{S}_d$, we find that our Hamiltonian can be written as
\begin{eqnarray}
H=H_A+H_J\\
H_J=-J_d\bf{S}_d\cdot\bf{S}\\
S_d^z = \frac {1} {2}(n_{d\uparrow} - n_{d\downarrow}),\hspace{.1in} &
S_d^+=a^{\dagger}_{d\uparrow}a_{d\downarrow} &
,\hspace{.1in}S_d^- = a^{\dagger}_{d\downarrow}a_{d\uparrow}.
\end{eqnarray}
The first question we answer with this model is, does the Kondo effect still
occur.
Before rigorous calculations are performed, a heuristic answer can be put forth
immediately.  Without loss of generality,
the defect energy can be taken to be $|\epsilon_d|\approx U>>|J_d|$, as $J_d$
is typically
a fraction of U.  The tunneling rate to the dot is determined by the matrix
element
$V_{kd}$.  Because this quantity is an adjustable parameter determined by the
width
of the tunnel junction connecting the dot to the source lead, we can set
$|J_d|>|V_{k_Fd}|$. This inequality is crucial in the analysis of what follows.
 For example, in the absence of the $J_d$ coupling, the form of the
antiferromagnetic interaction
that gives rise to the Kondo effect scales as $|J_K|\approx\frac {|V_{k_Fd}|^2}
{U}$. However,
in the limit that $|J_d|>|V_{k_Fd}|$, $|J_d|>>|J_K|$.  That is, the
exchange coupling exceeds the Kondo coupling and could hence ultimately
conspire to mask
the Kondo effect. Consider the case in which the d-level and the S-level are
singly
occupied. In the ferromagnetic case, $J_d>0$, the ground state of the dot is a
triplet. A Kondo effect should result in this case that is
determined by the total spin on the dot.  However, in the antiferromagnetic
case,
$J_d<0$, the ground state on the dot is a singlet.  As a consequence, there is
no
net spin to couple to the conduction electrons and the Kondo effect is
suppressed.
In the antiferromagnetic case, there must be an odd number of electrons on the
dot
for the Kondo effect to be observed.  This is the essential physics of this
model.

To prove the heuristic arguments given above, we diagonalize $H$ in the
subspace
of all singly-occupied states on the dot.  We first note that because
$[H_0,H_J]=0$, we can
work entirely with the eigenstates of the dot.  Let us define a generalized
eigenstate
$|q\rangle=|N;S_{tot},S_{tot}^z\rangle$, where N refers to the number of
electrons
on the d-level of the dot. Recall, we have set $S^z=\pm\frac {1} {2}$.  There
are
8 eigenstates in the dot basis:
$|0;\frac {1} {2}, S^z_{tot}=\pm\frac {1} {2}\rangle$ with total energy E=0
, the singlet $|1;0,0\rangle$ with energy $\epsilon_s=\epsilon_d+\frac {3J_d}
{4}$,
the triplets
$|1;1,\hspace{.05in}S^z_{tot}=0,\hspace{.05in}\pm 1\rangle$, with energy
$\epsilon_t=\epsilon_d-\frac {J_d} {4}$ and the doubly-occupied state,
$|2;\frac {1} {2},S^z_{tot}=\pm\frac {1} {2}\rangle$ with energy
$2\epsilon_d+U$. Each of the Fermion operators as well as the bi-linears such
as
 $\bf{S_d}\cdot\bf{S}$ can be expressed in terms of these 8 basis states.
Once this is done, matrix elements among these states can be determined
straightforwardly.  For an arbitrary wavefunction $|\psi\rangle$,
we are interested in solving the Schroedinger equation
\begin{equation}
\sum_p \langle q|H|p\rangle\langle p|\psi\rangle =E\langle q|\psi\rangle
\end{equation}
in the subspace of singly occupied states, N=1.  This is the relevant phase
space
for considering the Kondo effect.  To reduce the full $8\times 8$ to a $4\times
4$,
we rewrite the matrix elements involving the empty and doubly-occupied states
in terms of the singly occupied states.  The exact result is a $4\times 4$
Hamiltonian
matrix

\begin{equation}
\stackrel{\sim}{H} = \left( \begin{array}{llll}
   H_{s} & \displaystyle{
Q^{(-)}_{\uparrow\uparrow}-Q^{(-)}_{\downarrow\downarrow}} &\,\,
R_{\uparrow\downarrow} &
         -R_{\downarrow\uparrow} \\
   \displaystyle{Q^{(-)}_{\uparrow\uparrow}-Q^{(-)}_{\downarrow\downarrow}}
&\,\, H_{t,0} &
         -R_{\uparrow\downarrow} & -R_{\downarrow\uparrow} \\
    R_{\downarrow\uparrow} &\,\, -R_{\downarrow\uparrow} & H_{t,1} & 0\\
                           -R_{\uparrow\downarrow}   & -R_{\uparrow\downarrow}
& 0 & H_{t,-1}\\
                            \end{array} \right)\quad
      \begin{array}{l}
         \langle 1;0,0|\psi\rangle \\
         \langle 1;1,0|\psi\rangle \\
         \langle 1;1,1|\psi\rangle \\
         \langle 1;1,-1|\psi\rangle
      \end{array}
\end{equation}
where the singlet and triplet Hamiltonians are
\begin{eqnarray}
H_s = H_c + \epsilon_S + \sum_{\sigma}  Q^{(+)}_{\sigma\sigma}, \quad
H_{t,0} = H_c + \epsilon_t + \sum_{\sigma}  Q^{(+)}_{\sigma\sigma}\\
H_{t,1} = H_c + \epsilon_t + 2Q^{(+)}_{\downarrow\uparrow}, \quad H_{t,-1} =
H_c + \epsilon_t + 2Q^{(+)}_{\uparrow\downarrow}\\
\nonumber
\end{eqnarray}
and Q and R are the matrix elements
\begin{eqnarray}
Q^{(\pm)}_{\sigma,\sigma'} = \frac{1}{2} \sum_{kk'} V_{k'} V_k^{\ast} [\pm
(E + \epsilon_{k'} - H_c - 2\epsilon_d - U)^{-1} a^+_{k'\sigma} a_{k\sigma}
-(E - \epsilon_k - H_c)^{-1} a^+_{k'\sigma'} a_{k\sigma'}]\\
R_{\sigma\sigma^{'}} = \frac{1}{\sqrt{2}} \sum_{kk'} V_{k'} V_k^{\ast} [(E +
\epsilon_{k'} - H_c - 2\epsilon_d - U)^{-1} + (E - \epsilon_k - H_c)^{-1}]
a^+_{k'\sigma} a_{k\sigma^{'}}\\
\nonumber
\end{eqnarray}

The matrix elements $R$ and $Q$ contain all powers of the coupling to the
conduction
electrons.  To lowest order, they scale roughly as $\frac {|V_k|^2} {U}$.  The
Hamiltonian matrix
can be partitioned into $1\times 1$ singlet and $3\times 3$ triplet subspaces
provided that
the differences between the diagonal elements exceeds the off-diagonal matrix
elements
$R$ and $Q$.  The diagonal elements differ by the spin-exchange $J_d$.
Consequently,
the partitioning into singlet and triplet subspaces is valid provided that
$|J_d|>\frac {|V_k|^2} {U}$.  The effective Hamiltonian in each subspace that
is
 valid to second order in the coupling to the leads can be obtained by setting
$E= \epsilon_s+H_c$ and $E=\epsilon_t+H_c$ in the denominators of $R$ and $Q$
and transcribing
the basis state representation back to the original Fermion operators.
In the ferromagnetic case $(J_d > 0)$, the reduced Hamiltonian in the triplet
subspace
is
\begin{equation}
H_{eff}^{triplet} = H_c +\epsilon_t+\displaystyle{ \sum _{kk'\sigma}
W^t_{kk'}(\psi ^{\dagger}_{k'} \psi _k)}
- \displaystyle{ \sum  _{kk'}J_{kk'}^t(\psi^{\dagger}_{k'} {\bf\sigma} \psi
_k)}
 \cdot {\bf S}
_{tot}
\end{equation}
where $H_c$ is the Hamiltonian for the free conduction electrons, $\bf{\sigma}$
is
the Pauli spin matrix, $\psi_k$ is the two-component spinor
\[\psi_k= \left( \begin{array}{c}
a_{k\uparrow}\\
a_{k\downarrow}
\end{array} \right) \]
the antiferromagnetic coupling constant is
\begin{equation}
J_{kk'}^t=2V_{k'd}V^*_{kd}\left( \frac {1}
{\epsilon_{k'}-2\epsilon_d-U+\epsilon_t}
 - \frac {1} {\epsilon_k-\epsilon_t}\right)
\end{equation}
and
\begin{equation}
W_{kk'}^t=\frac{1} {2}V_{k'd}V^*_{kd}\left( \frac {1}
{\epsilon_{k'}-2\epsilon_d-U+\epsilon_t}
 + \frac {1} {\epsilon_k-\epsilon_t}\right).
\end{equation}
That $J^t_{kk'}$ is negative can be seen immediately because the largest energy
scale
in the denominator is $U$ which enters with a $-$ sign.  The spin interaction
obtained
in this limit is identical to the usual Kondo coupling except in this case the
total
spin on the dot enters.  For a triplet state $S_{tot}=1$.  Consequently,
ferromagnetic
exchange gives rise to a S=1 Kondo problem.  The S=1 Kondo problem is an
example of an
undercompensated spin problem in which
the conduction electrons only partially screen the spins on the dot. The
remaining
unscreened spin couples ferromagnetically to the conduction band.  The only
qualitative difference between the S=1 Kondo
problem and S=$\frac{1} {2}$ is the behavior of the magnetic susceptibility.
As a result
of the undercompensation, the susceptibility does not vanish at T=0 in the S=1
problem.

Consider now the more experimentally-relevant antiferromagnetic
case\cite{foot}.  We recover in this limit an effective
Hamiltonian of the form
\begin{equation}
H^{singlet}_{eff} = H_c + \epsilon_s +\displaystyle{
\sum_{kk'}} W^s_{kk'}(\psi ^{\dagger}_{k'}
\psi _k) + \frac{1}{8J_d} \displaystyle{ \sum _{\stackrel{k_1 k_1 '}{k_2
k_2 '}} J^s_{k_1 k_1 '}
J^s_{k_2 k_2 '}(\psi^{\dagger}_{k_1 '} \bf{\sigma} \psi _{k_1})} \cdot (\psi
^{\dagger}_{k_2 '}
\bf{\sigma} \psi_{k_2})
\end{equation}
where $W^s_{kk'}$ and $J^s_{kk'}$ are identical to their triplet counterparts
with
$\epsilon_s$ replaced by $\epsilon_t$.  As is evident the spin coupling only
involves
the conduction electrons and is $O((V_{kd})^4)$. Further, the overall sign of
this interaction is negative or
ferromagnetic.  Consequently, there is no antiferromagnetic exchange
interaction
that can produce a Kondo effect to fourth order in the coupling to the band
electrons.
The physical origin of the absence of the Kondo effect here is the stability
of the singlet on an energy scale $J_d$. As a result, the Kondo coupling
constant
must be cut off at this energy scale.  Consequently, it cannot diverge and give
rise
to a bound state at the dot.  This result is consistent with the heuristic
arguments
of Ng and Lee\cite{nglee} on the role of spin exchange in a quantum dot and a
mean-field %N->\infty$
limit of the 2-impurity Anderson model in the presence of
spin-exchange\cite{millis}.
Jones, Kotliar and Millis\cite{millis} found that in the  $N->\infty$ limit of
this model, a phase transition
occured which suppressed the Kondo effect if the bare exchange interaction
exceeded
a critical value.  The critical condition is similar to the one used here,
namely
$|J_d|>\frac {|V_{kd}|^2} {U}$.  There still remains one chance for the Kondo
effect to be observed when $J_d<0$.
If the number of electrons on the dot is odd, or equivalently we restrict
ourselves to
the N=0 subspace, the standard S=$\frac{1} {2}$ Kondo
problem is recovered.  If experiments are going to detect the Kondo effect, the
total
number of electrons on the dot must be carefully controlled.

In deriving the results in the Kondo regime, we have performed
2nd-order perturbation theory in the coupling to the leads.  It is possible to
construct a Schriefer-Wolff- type\cite{swolff} transformation that eliminates
the coupling to the
leads.  The result
\begin{eqnarray}
\hat S=\frac{1} {2S+1} \sum_{k,\sigma,\sigma'} V_k \left\{ \left[ \left(\frac
{S+1} {E_t^{(-)}} +
\frac {S} {E_s^{(-)}}\right)(1-n_{d-\sigma})+\left(\frac {S+1} {E_t^{(+)}} +
\frac {S} {E_s^{(+)}}\right)n_{d-\sigma}\right] \delta_{\sigma\sigma'}\right.
\nonumber\\
\left. +2({\bf S} \cdot {\bf s_{\sigma\sigma'}})\left[ \left(\frac {1}
{E_t^{(-)}} -
\frac {1} {E_s^{(-)}}\right)(1-n_{d-\sigma'})+\left(\frac {1} {E_s^{(+)}} -
\frac{1} {E_t^{(+)}}\right)n_{d-\sigma'}\right] \right\}
c_{k\sigma}^{\dagger}d_{\sigma'}\nonumber\\
- h.c.\\
\nonumber
\end{eqnarray}
can be used to derive the effective Hamiltonians in the singlet and triplet
subspaces.
In the above, $E^{(-)}_{s,t}=\epsilon_k - \epsilon_{s,t}$ and
$E^{(+)}_{s,t}=\epsilon_k -2\epsilon_d -U +\epsilon_{s,t}$. In the singlet
subspace,
$S=0$ and only those virtual transitions involving the triplet state survive.
This transformation
successfully eliminates the coupling to the leads in the limit $\frac {|V_k|^2}
{U}<<1$ and hence
is consistent with the perturbative treatment developed here.

We now calculate the conductance at finite temperature in the presence of
spin-exchange.
To facilitate this we need the average occupancy on the dot $\langle
n_{d\sigma}\rangle$.
This quantity is obtained by integrating the imaginary part of the  d-electron
Green function,
$G_{d\sigma}(\omega)=\langle\langle a_{d\sigma};
a_{d\sigma}^{\dagger}\rangle\rangle$, weighted with
the Fermi-Dirac distribution function.  Standard equations of motion methods
\cite{eom}\cite{lacroix} can
 be used to formulate an accurate expression for $G_{d\sigma}(\omega)$.  Each
level of iteration generates
a new heirarchy of Green functions for which new equations of motion must be
derived. To illustrate,
the Heisenberg equations of motion for $G_{d\sigma}(\omega)$ generate two new
Green functions,
$\langle\langle n_{d-\sigma}a_\sigma;a_{d\sigma}^{\dagger}\rangle\rangle$ and
$\langle\langle {\bf S}\cdot{\bf
S_d}a_{d\sigma};a_{d\sigma}^{\dagger}\rangle\rangle$
Equations of motion for this set of Green functions as well as for the new
Green functions
that appear at this level were derived and solved self-consistently for the
impurity
density of states by invoking the Hartree-Fock closure. The density of states
obtained
at the 3rd level of iteration is sufficient to describe the Kondo effect. As we
have already described the $T=0$ phase, we focus on the
experimentally-accessible high-temperature
limit. The conductance
\begin{equation}
\Gamma=\frac {-2e^2} {h} \frac {\Delta} { k_BT}\int
f_{FD}(\omega)(1-f_{FD}(\omega))
Im G_{d\sigma}(\omega+0)d\omega
\end{equation}
was calculated using the standard Landauer formula\cite{lee}\cite{landauer}. In
Eq.(16),
$\Delta=\frac {-1} {\pi}\sum_k |V_k|^2\delta (\omega-\epsilon_k)$ To illustrate
the role of spin-exchange, we report here the
infinite U limit of $G_d(\omega)$ at the second level equations of motion. We
find
that the second level Green function
\begin{equation}
G_d(\omega)= \frac {\frac {3} {4}\left( 1-\langle n_{d-\sigma}\rangle\right) +
\frac {1} {2}\langle {\bf S}\cdot{\bf S_d}\rangle} {\omega-\epsilon_s-\Sigma_0
\left( 1-\langle n_{d-\sigma}\rangle\right)} + \frac {\frac {1} {4}\left
( 1-\langle n_{d-\sigma}\rangle\right) -
\frac {1} {2}\langle {\bf S}\cdot{\bf S_d}\rangle} {\omega-\epsilon_t-\Sigma_0
\left( 1-\langle n_{d-\sigma}\rangle\right)}
\end{equation}
contains a contribution for the singlet and triplet states with differing
spectral weights.
In Eq. (18), the self energy is $\Sigma_0=\sum_k |V_k|^2
(\omega-\epsilon_k)^{-1}\approx -i\Delta$.
This expression clearly illustrates that the singlet (first term) and the
triplet
(second term) spectral weights differ.  We expect that the conductance into the
singlet and triplet levels
should reflect the asymmetry in the spectral weights.  The conductance
calculated from
Eq. (17) (with the 3rd level Green function) is shown in Figure 1 as a function
of the chemical potential.  Illustrated
clearly is the asymmetry in the conductance peaks centered at $\epsilon_s$ (the
first
peak) and at $\epsilon_t$ (2nd peak).  In figure 1, $J_d=-.1U$.  Hence, the
singlet
is the ground state.  In the absence of $J_d$, the singlet and triplet peaks
would
coalesce into a single peak as in the standard Anderson model.  The higher
peaks
in the conductance occur at energies $2\epsilon_d-\epsilon_t+U$ and
$2\epsilon_d-\epsilon_s+_U$, respectively. The upper peaks appear inverted
because
for $J_d<0$, $\epsilon_s<\epsilon_t$.
In the ferromagnetic regime, the triplet peak dominates
and it is the neighbouring singlet states that lead to the asymmetry in the
peak heights in the
conductance as illustrated in Figure 2.  We conclude then that spin exchange in
zero magnetic field leads
to peak height alternation in the conductance that is identical in form to the
experimental\cite{kastner}
trends seen in the presence of a magnetic field.  Ultimately, the ferromagnetic
and
antiferromagnetic cases can be distinguished by a low temperature study of the
Kondo
phase.

\acknowledgments
We thank Patrick Lee and Yigal Meir for insightful remarks. This work is
supported in part by the NSF and the donors of the Petroleum Research
Fund of the American Chemical Society, and the Director of Energy Research,
Office of Basic
Eerngy Sciences. Ames Laboratory is operated for the U. S. DOE by Iowa State
University
under Contract No. W-7405-ENG 82.

\newpage
\centerline{\bf Figure Captions}

\bigskip

\noindent Figure 1: Conductance (measured in units of $e^2 \over h$) as a
function
of the chemical potential (measured in units of $U$) as computed from Eq. (17)
using the third-level
decoupling of the equations of motion for the Green function for $J_d=-0.1U$
 and $\Delta=0.001U$. Each set of two
peaks corresponds to a singlet and triplet pair, the singlet being lower in
energy in the antiferromagnetic case.

\noindent Figure 2: Same as Figure 1 but for the ferromagnetic coupling,
$J_d=0.1U$.
Each set of two
peaks corresponds to a singlet and triplet pair, the triplet being lower in
energy in the ferromagnetic case.

\end{document}